\documentclass[journal,transmag]{IEEEtran}

\ifCLASSINFOpdf

\else

\fi

\usepackage{graphicx}
\usepackage{epstopdf}
\usepackage{subfig}

\begin{document}

\title{A Random Matrix Theoretical Approach to Early Event Detection Using Experimental Data}

\author{\IEEEauthorblockN{Yingshuang Cao\IEEEauthorrefmark{1},
Long Cai\IEEEauthorrefmark{1},
Robert C. Qiu\IEEEauthorrefmark{1,2},
Jie Gu\IEEEauthorrefmark{1},Xing He\IEEEauthorrefmark{1},Qian Ai\IEEEauthorrefmark{1}, and
Zhijian Jin\IEEEauthorrefmark{1}}}

\markboth{Journal of \LaTeX\ Class Files,~Vol.~13, No.~9, September~2014}%
{Shell \MakeLowercase{\textit{et al.}}: Bare Demo of IEEEtran.cls for Journals}

\IEEEtitleabstractindextext{%
\begin{abstract}
In this paper, High-dimensional data analysis methods are proposed to deal with random matrix which is composed by the real data from power network before and after the fault. The mean  spectral radius (MSR) of non-Hermitian random matrices is defined as a statistic analytic for the fault detection. By analyzing the characteristics of random matrices and observing the changes of the spectral radius of random matrices, grid failure detection will be achieved. This paper describes the basic mathematical theory of this big data method, and the real-world data of a certain China power grid is used to verify the methods.
\end{abstract}

\begin{IEEEkeywords}
Fault recognition and diagnosis, Big Data, spectral radius, random matrices and modern smart grid
\end{IEEEkeywords}}

\maketitle

\IEEEdisplaynontitleabstractindextext

\IEEEpeerreviewmaketitle

\section{Introduction}

\IEEEPARstart{T}{HE} increasing data have become a strategic resource in smart grids.[1] These datasets contains valuable information and signals. Our team uses the theory of big data to detect fault through analysing these datasets. The result can be used to improve the safety and stability of modern power system.

The amount of real-time information in power system increases quickly with the development of the smart grid. These data are mostly unstructured and come from different resources. In recent years, increasing data are collected from Phasor Measurement Units (PMUs), Intelligent Electronic Devices (IEDs), Supervisory Control and Data Acquisition (SCADA) and so on [2]. There is a complex relationship between these high-dimensional data. The current research mainly focused on determining fault detection signal and the detection signal treatment [3-4]. But when applied to complex power grid, these methods become inefficient and invalid. With the advent of the era of Big Data, the theory of big data has been applied to many fields [5-7]. It has been proven to be a good method to analyze these massive and high-dimensional data. In recent years, Big Data are also studied in power system analysis of faults and disturbances and some promising researches were made [8,9].

Fault detection is the keystone of our research and also the key to improve the safety and stability of modern power system. Fault detection requires continuous monitoring and processing of massive quantities of data in order to detect and identify emerging patterns, which means telling signals from noises. For smart grids, uncertain locations of PMUs, small random loads/generators fluctuations and sample errors could be regarded as noises, whereas sudden changes of loads/generators, system faults, network reconfiguration as signals. It is a big challenge to respond within a tolerable elapsed time or hardware resources by extracting analysis from streaming data in the smart grids.

The work of [8] is the most related work to our paper. Xie, Chen and Kumar used the principal component analysis (PCA) for early event detection with both simulations and experimental data. Our work is built upon their work but different from their work in a fundamental sense. PCA is a widely adopted approach in unsupervised machine learning. First, they used the features extracted from the data in the training phase; second, they used the extracted features for early event detection. Their pioneering work motivates our research on one hand; we are motivated by the line of research in the spirit of [5-7,10], on the other hand. The central idea of our approach is to model a large power system as a high-dimensional statistical problem. So we can exploit the high-dimensionality of the massive datasets. The high-dimensionality of the data is a blessing, not a curse. The work of [13] is also related here. We only study the power grid using simulations in that paper. This current paper is regarding experimental data. The objectives of two papers are also different.

In this paper, random matrix theory is used to model the real-life data from a certain power grid in China. This is essentially an anomaly detection problem in the literature of big data. It may be modeled as binary hypothesis testing: normal hypothesis $ {\cal H}_0 $ (no signal present) and anomaly hypothesis $ {\cal H}_1 $ (signal present). Traditionally, we model the dataset in hypothesis $ {\cal H}_1 $ as a standard Gaussian noise random vector whose entries are $ X_i \sim {\cal CN} (0,1),\rm{ } i=1, \cdots ,n $. In our real-life data, the traditional Gaussian model seems be not natural. We use the universality of random matrix theory to argue that the Gaussian random matrix can be used to model our real data, even though the data are non-Gaussian. This universality is only valid when the data dimension is high. This is in some sense like the central limit theorem.

The anomaly hypothesis $ {\cal H} _1 $ is claimed when the data deviates from the normal hypothesis $ {\cal H} _0 $. Our central analysis is focused on the latter.

We use the measurement data without the knowledge of any other power grid information. In other words, our black box approach is model-free and purely data-driven, relying on the high dimensionality of the data. Abstract statistical laws in high dimension probability serves as the basis for our approach [5,6,7]. The unified feature of the abstract results have proved effective using real-life data.

This paper is organized as follows. Section II is used to explain, principal component analysis and random matrix theory. Based on the theories in Section II, related algorithms and test sample are proposed in Section III. Conclusions and possible future research directions are presented in Section IV.

\section{The Basic Principle of Random Matrix Theory And Principle Component Analysis}

\subsection{Random Matrix Theory}


\subsubsection{Marchenko-Pastur Law (MP Law)}

The MP Law describes the asymptotic behavior of singular values of large rectangular random matrices [11]. Let $X=\left\{ {{\xi _{ij}}} \right\}$ be a $N \times T\left( {N/T = c \in \left( {0,1} \right)} \right)$ random matrix whose entries are independent identically distributed (i.i.d.) random variables with mean $\mu  = 0$ and variance ${\sigma ^2} < \infty$.

The empirical spectrum density (ESD) of the corresponding sample covariance matrix $S = {\textstyle{1 \over T}}XX^H$ ( i.e. $f\left( {\lambda \left( S \right)} \right)$ ) converges to the distribution of MP Law with density function:

\begin{equation}
{f_{MP}}(x) = \left\{ \begin{array}{ll}
{\textstyle{1 \over {2\pi xc{\sigma ^{\rm{2}}}}}}\sqrt {(b - x)(x - a)} {\rm{  }}&,{\rm{ a}} \le {\rm{x}} \le {\rm{b}}\\
0&,{\rm{ otherwise}}
\end{array} \right.
\end{equation}

where $a={\sigma ^2}{(1 - \sqrt c )^2},\rm{ }b = {\sigma ^2}{(1 + \sqrt c )^2},\rm{ }c = N/T$.

\subsubsection{Kernel Density Estimation (KDE)}

A nonparametric estimate of the empirical spectral density of the sample covariance matrix is used

\begin{equation}
{f_n}(x) = {\textstyle{1 \over {nh}}}\sum\nolimits_{i = 1}^n {K({\textstyle{{x - {\lambda _i}} \over h}})}
\end{equation}

where $\lambda_i(i=1,2,\cdots,n)$ are the eigenvalues of $S$, and $K\left(  \cdot  \right)$ is the kernel function for bandwidth parameter $h$.

\subsubsection{ The Single-Ring Law}

For each $n \ge 1$, let $A_n$ be a random matrix which admits the decomposition[12]:

\begin{equation}
{A_n} = {U_n}{T_n}{V_n},\quad {T_n} = {\mathop{\rm diag}\nolimits} ({s_1}, \cdots ,{s_n})
\end{equation}

where $S_i$ are positive, and $U_n$ and $V_n$ are two independent random unitary matrices which are Haar-distributed independently from $T_n$. In probability, the ESD of $A_n$ converges weakly to a deterministic measure whose support is under certain mild conditions. Some outliers to single ring law are observed.

Consider the matrices product $\widetilde Z = \prod\nolimits_{i = 1}^L {{X_{u,i}}} $, where $X_u$ is the singular value equivalent of the rectangular $N \times T$ non-Hermitian random matrix $\widetilde X$, whose entries are independent identically distributed (i.i.d.) variables with mean $\mu ({\widetilde x_{k,:}}) = 0$ and variance ${\sigma ^2}({\widetilde x_{k,:}}) = 1$ for $k=(1,2,\cdots,N)$. The matrices product $\widetilde Z$ is converted to $Z$ by a transform which make the variance to ${\sigma ^2}({z_{:,k}}) = 1/N$ for $k=(1,2,\cdots,N)$. Thus, the empirical spectrum density of $Z$ converges almost surely to the same limit given by

\begin{equation}
{f_Z}(x) = \left\{ \begin{array}{ll}
{\textstyle{1 \over {\pi c\sigma }}}{\left| \lambda  \right|^{(2/\alpha  - 2)}}{\rm{  }} &,{{(1 - c)}^{\alpha /2}} \le \left| \lambda  \right| \le 1\\
0 &,{\rm{otherwise}}
\end{array} \right.
\end{equation}

as $N,T \to \infty $ with the ratio $N/T = c \in \left( {0,{\rm{ }}1} \right]$. On the complex plane of the eigenvalues, the inner circle radius is ${(1 - c)^{\alpha /2}}$ and outer circle radius is unity [13]. Moreover, $S=ZZ_H$ is able to acquired and its ESD converges to the distribution of MP Law.

\subsubsection{Universality of the MP Law}

Akin to the central limit theorem, universality [5, page 347] refers to the phenomenon that the asymptotic distributions of various covariance matrices (such of eigenvalues and eigenvectors) are identical to those of Gaussian covariance matrices. These results let us calculate the exact asymptotic distributions of various test statistics without restrictive distributional assumptions of matrix entries. The presence of the universality property suggests that high-dimensional phenomenon is robust to the precise details of the model ingredients [14]. For example, one can perform various hypothesis tests under the assumption that the matrix entries not Gaussian distributed but use the same test statistic as in the Gaussian case.

The data of power grid below can be viewed as a spatial and temporal sampling of the random graph. Randomness is introduced by the uncertainty of spatial locations and the system uncertainty. Under real-life applications, we cannot expect the matrix entries follow i.i.d. distribution. Numerous studies based on both simulations [10] and experiments however, demonstrate that the MP law is universally followed. In such cases, universality properties provide a crucial tool to reduce the proofs of general results to those in a tractable special case---the i.i.d. case in our paper.

\subsection{Principal Component Analysis (PCA)}

A method based on PCA is proposed to testify our result. PCA is the most commonly used linear dimensionality reduction methods. PCA reduces the dimensionality by preserving the most variance of original data. Due to the increasing size of PMU data, the dimensionality analysis of PMU data has been studied in recent literature. The analysis shows the rank of PMU data matrix is low. The massive PMU data essentially lie in a much reduced dimensional space.

Let $P$ denote the number of available PMUs across the whole power network, $p$ denote the number of PMUs used in computation. Each PMU provides $l$ measurements. There are a number of PMUs in interconnected power network, with each PMU providing measurements of voltage magnitude and power flow. At each time sample, a total of $p \times l$ measurements are collected. In this paper, we conduct the analysis for each category of measurements independently. Define the measurement vector $\overrightarrow y = {[{y^{(1)}},{y^{(2)}}, \cdots ,{y^{(p)}}]^T}$ containing the $p$ measurements. Use to denote the measurements of time t. Define the measurement matrix $ Y_{p \times T}^{i + T} = [{\overrightarrow y _{i + 1}},{\overrightarrow y _{i + 2}}, \cdots ,{\overrightarrow y _{i + T}}] $,
The PCA-based event detection at time $i+T+1$ is described as follows [8]:

1) Calculate the matrix of $ {C_Y} = Y_{p \times T}^{i + T}{\left( {Y_{p \times T}^{i + T}} \right)^T} $.

2) Calculate the $T$ nonzero eigenvalues and vectors of $C_Y$.

3) Rearrange the $T$ eigenvalues in decreasing order, with the eigenvectors being the principal components (PCs).

4) Out of the $N$ PCs, select the $m$ largest eigenvalues and corresponding PCs.

5) Form a new m-dimensional subspace from the $m$ PCs.

6) Project the vector ${\overrightarrow y _{i + T + 1}}$ onto the m-dimensional PC-based space.

7) Calculate the length of the projection vector.

\section{Numerical Examples}

This case is based on the field data. The operation data of a certain power grid in China is collected. It includes the data of substation, breaker, line, bus, generator, frequency and so on. All of the PMU data are real-life data from five different interconnected power grids in China. Fig.1 shows the structure of the data lists from our test power grid. The sampling interval is 1 min (s) and the sampling duration is 4320mins (3days).

These data are exported from the power grid database and the time of grid failure is unknown in advance. After our analysis, the result has been compared with the real-life situation. At first the data of power flow are analyzed by PCA method, but valid information isn't got by this way (Fig.2).
Then the output data of active power of all generators are placed in a random matrix in time series. For 10 samples (10mins), we constitute a single random matrix and normalize each row of the matrix and calculate the mean spectral radius of every matrix. The mean spectral radius is regarded as a big data analytic index changes over time.

\begin{figure}[htbp]
  \centering
  \includegraphics[width=0.5\textwidth]{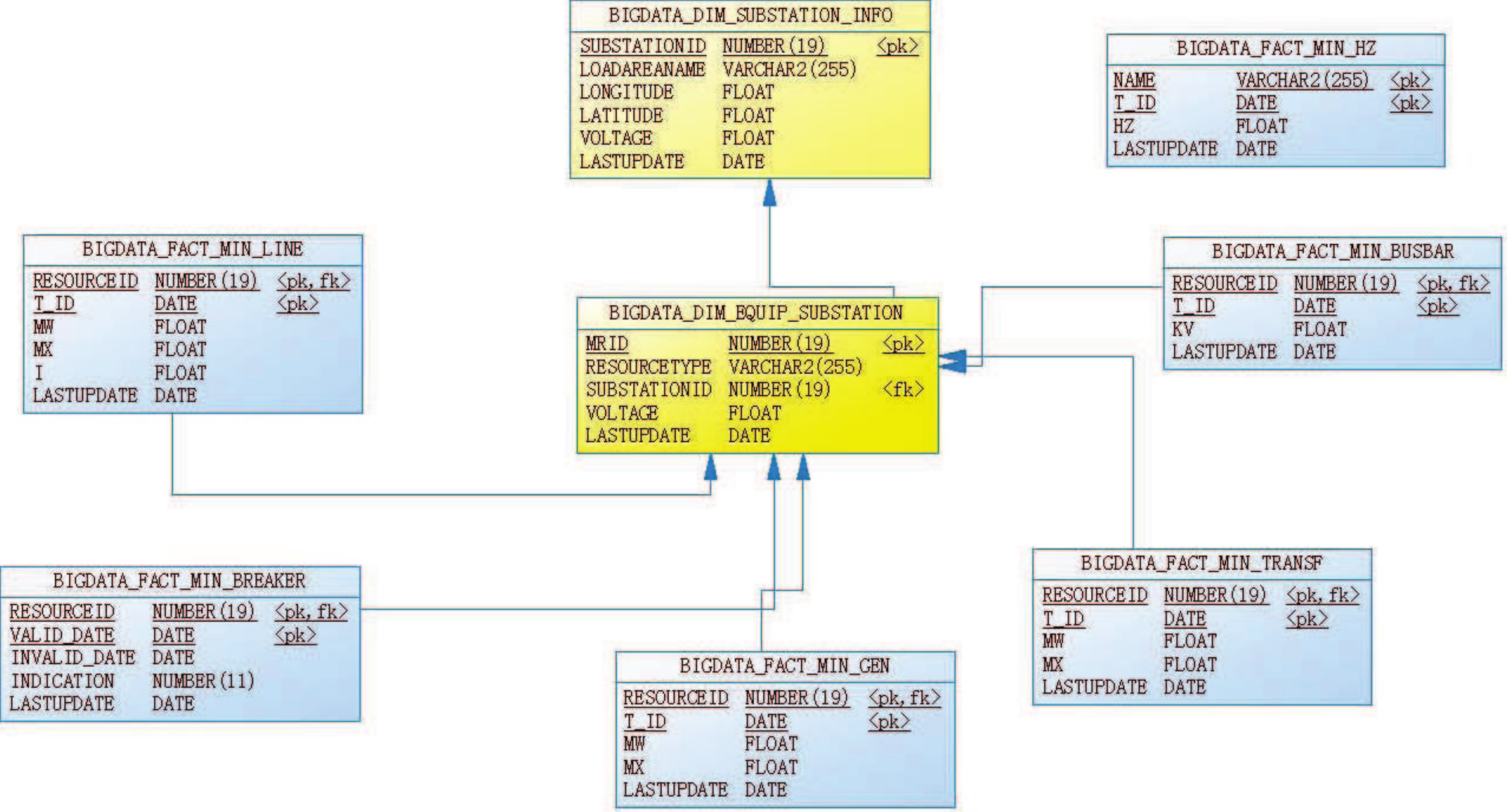}
  \caption{The structure of the data lists from a certain power grid in China}\label{fig:1}
\end{figure}

\begin{figure}[htbp]
  \centering
  \includegraphics[width=0.5\textwidth]{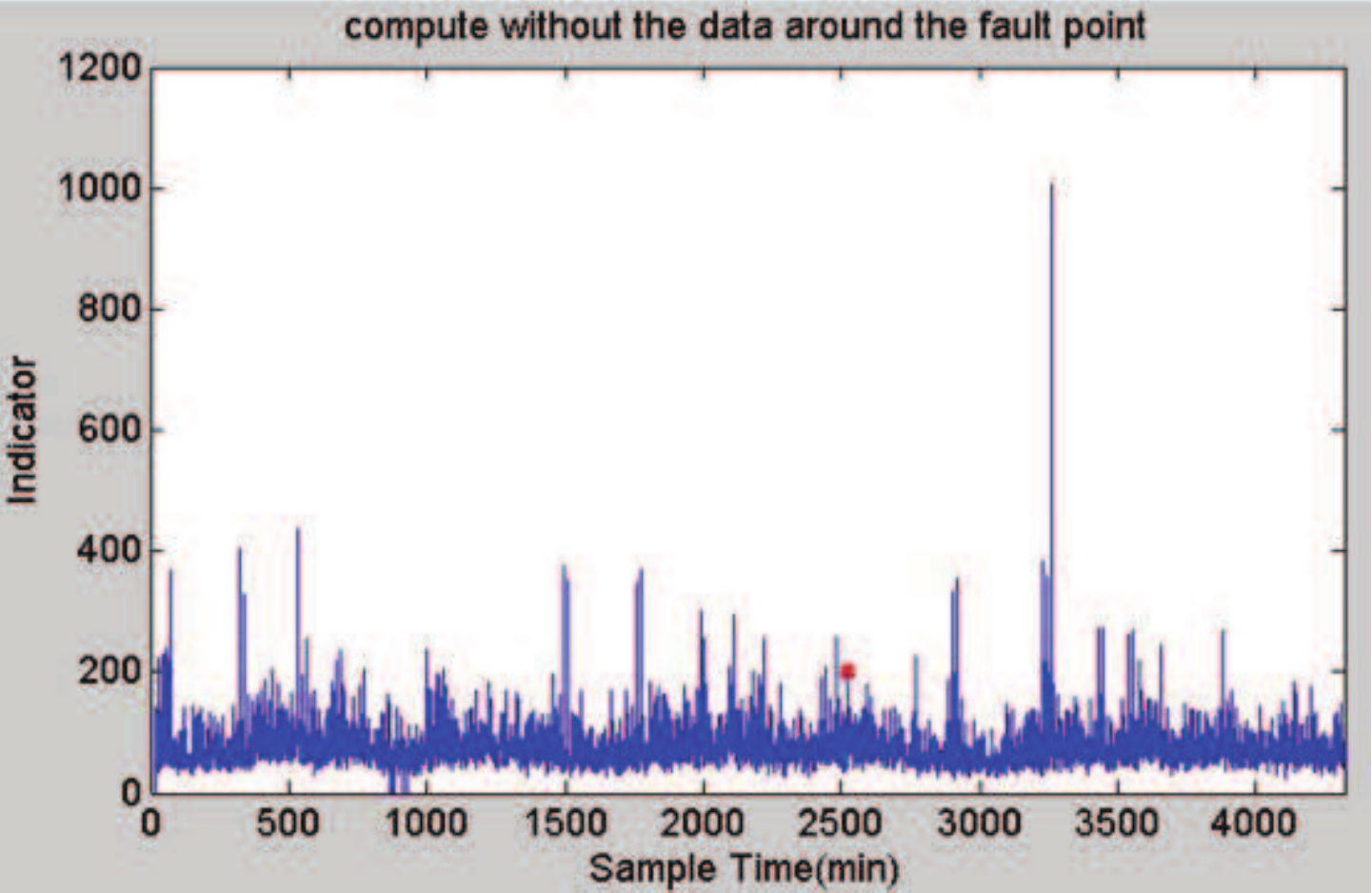}
  \caption{The results of load flow analysis}\label{fig:2}
\end{figure}

\begin{figure}[htbp]

\centering
\subfloat[Prefault]{
\label{fig:improved_subfig_a}
\begin{minipage}[t]{0.5\textwidth}
\centering
\includegraphics[width=\textwidth]{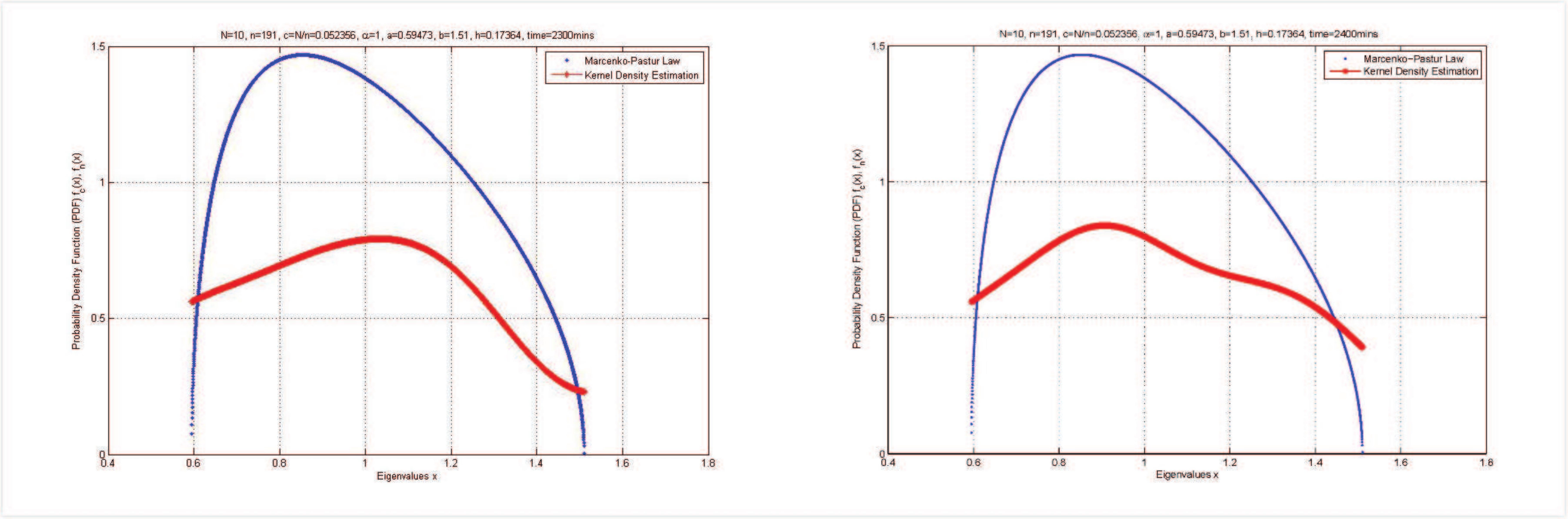}
\end{minipage}}

\subfloat[fault]{
\label{fig:improved_subfig_a}
\begin{minipage}[t]{0.5\textwidth}
\centering
\includegraphics[width=\textwidth]{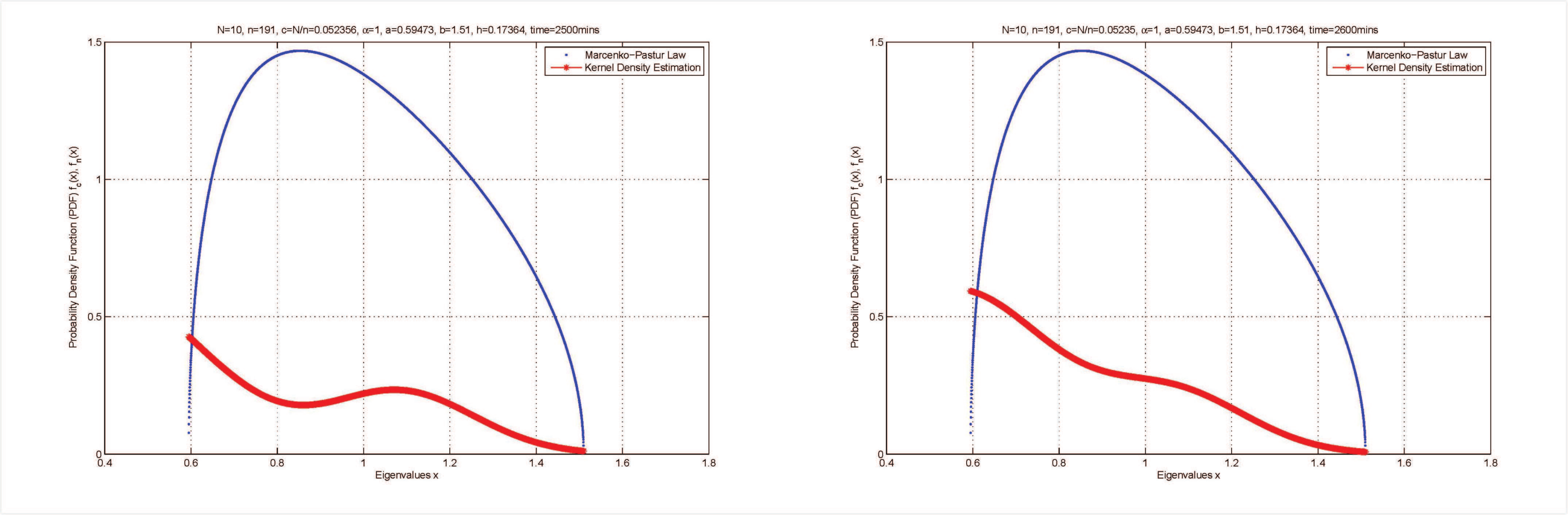}
\end{minipage}}

\subfloat[After fault]{
\label{fig:improved_subfig_a}
\begin{minipage}[t]{0.5\textwidth}
\centering
\includegraphics[width=\textwidth]{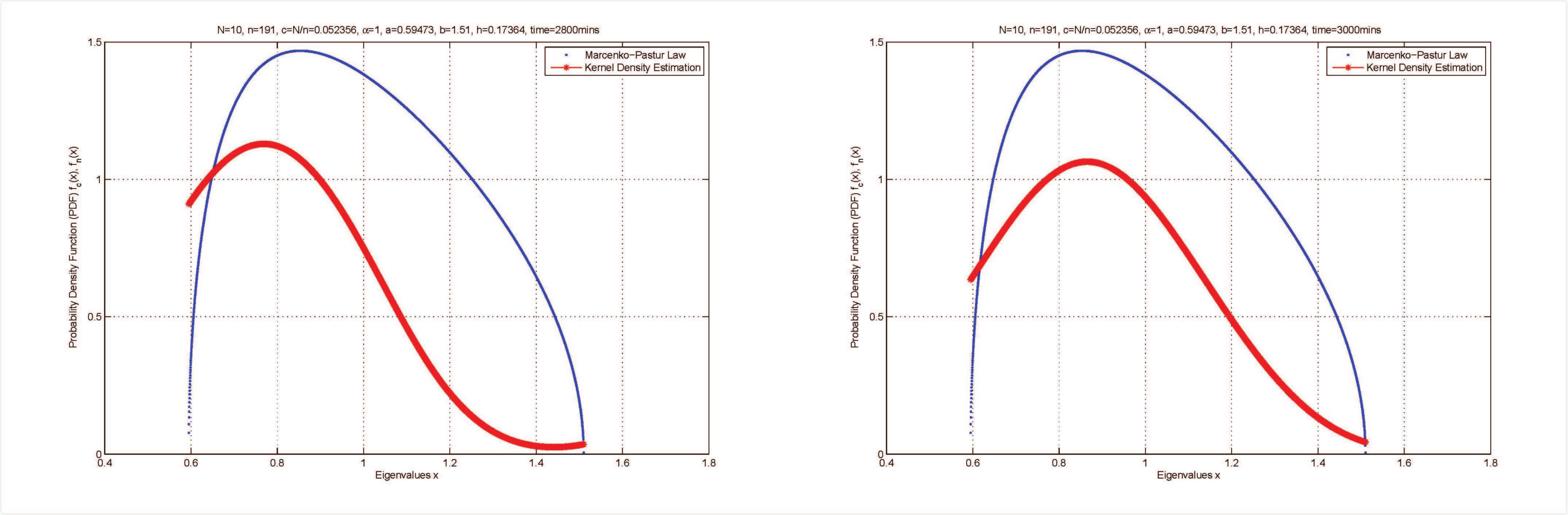}
\end{minipage}}

\caption{Empirical eigenvalue density functions of random matrices. The Marchenko-Pastur law is given in eq.(1)}
\end{figure}

Fig. 3 shows the relationship between Marcenko-Pastur Law and Kernel Density Estimation. Their matching degree will decrease when the grid fails. At the same time, the blue curve represents the M-P Law and the red curve represents the empirical eigenvalue density. When the degree of coincidence of the blue and red lines is higher, the matrix lines with the M-P Law better. Compared (a) with (b), the empirical eigenvalue density of the pre-fault matrix lines with the M-P Law much better than the fault matrix. In such a system, loads dramatic change or system failures  break the random distribution of system data, i.e. i.i.d. Under statistics perspective, it leads to a fluctuation in a certain direction, as well as the deviation of M-P law for the system data. When  other generators increase their output to maintain the stability of the grid, the random matrix will meet the i.i.d. gradually, so the blue and red lines close to each other again in (c). From a statistical perspective, the pre-fault describes the global behavior (the M-P law) of the power grid, while the fault tells the local behavior of this grid. We know by experience that the fault tends to occur locally.   The M-P law is the asymptotic distribution when the data size grows to infinity.

\begin{figure}[htbp]
  \centering
  \includegraphics[width=0.5\textwidth]{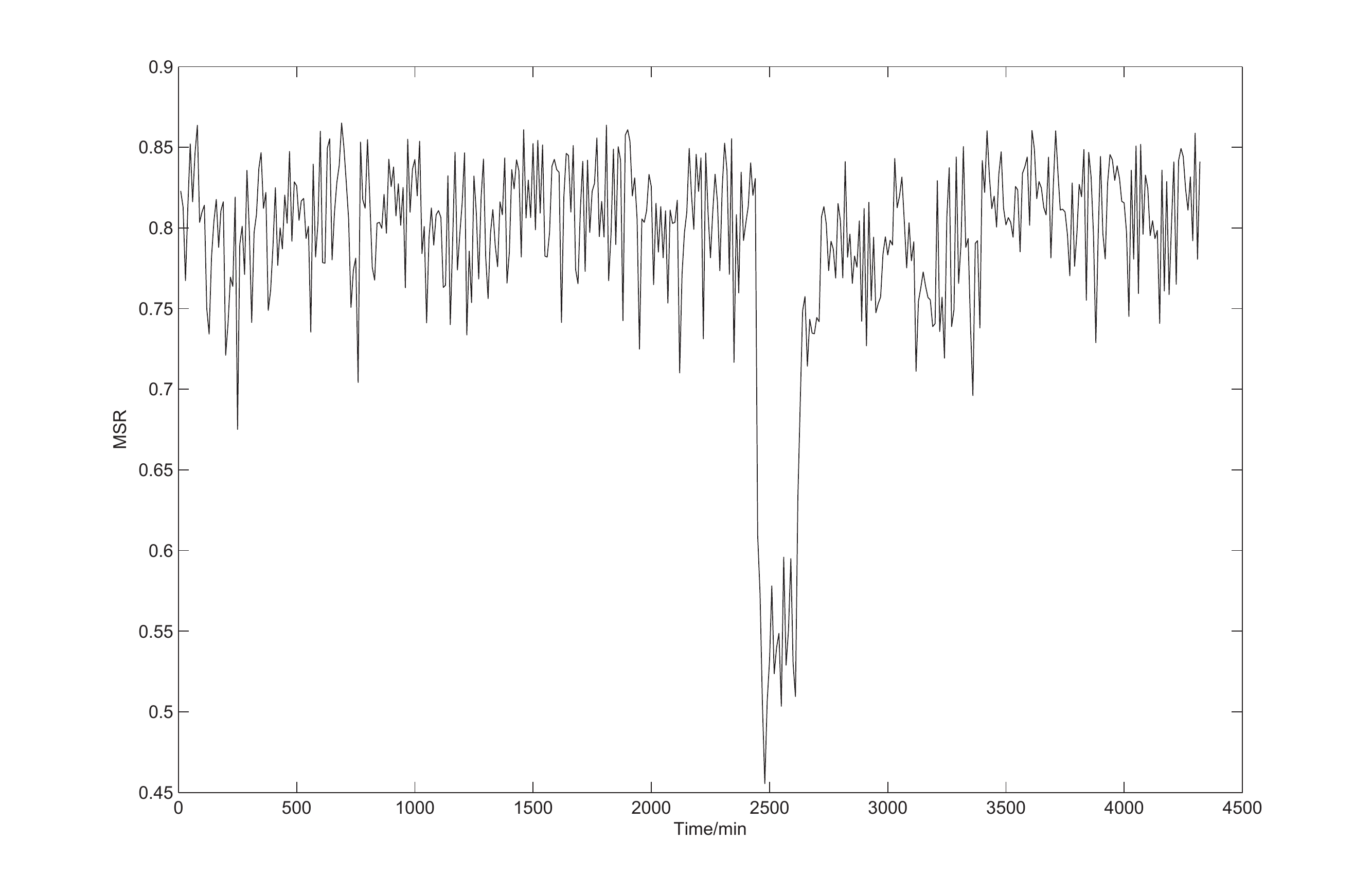}
  \caption{The average radius using time series for the events}\label{fig:4}
\end{figure}

\begin{figure}[htbp]
  \centering
  \includegraphics[width=0.4\textwidth]{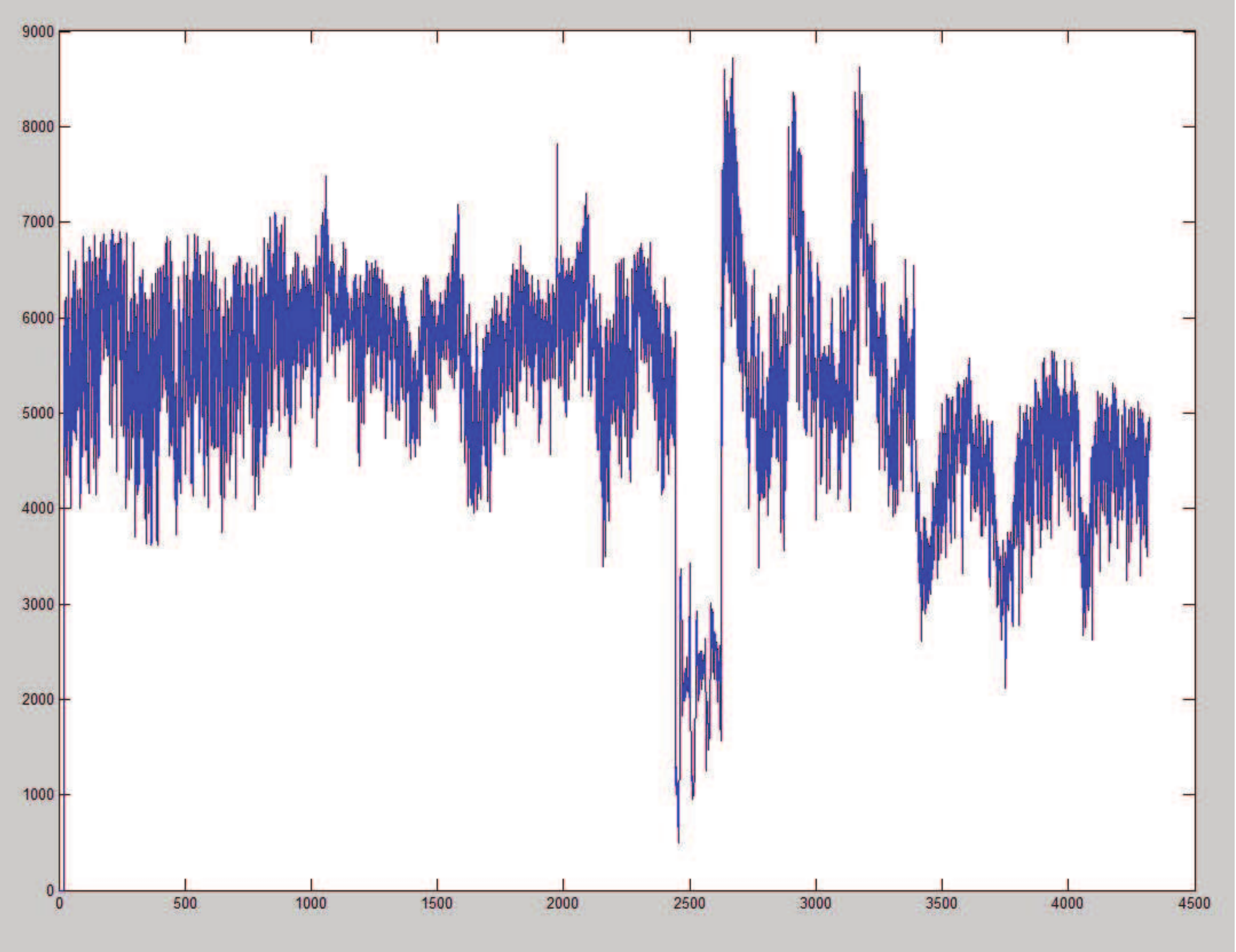}
  \caption{Generator unit output analysis chart by PCA for the events}\label{fig:5}
\end{figure}

\begin{figure}[htbp]
  \centering
  \includegraphics[width=0.5\textwidth]{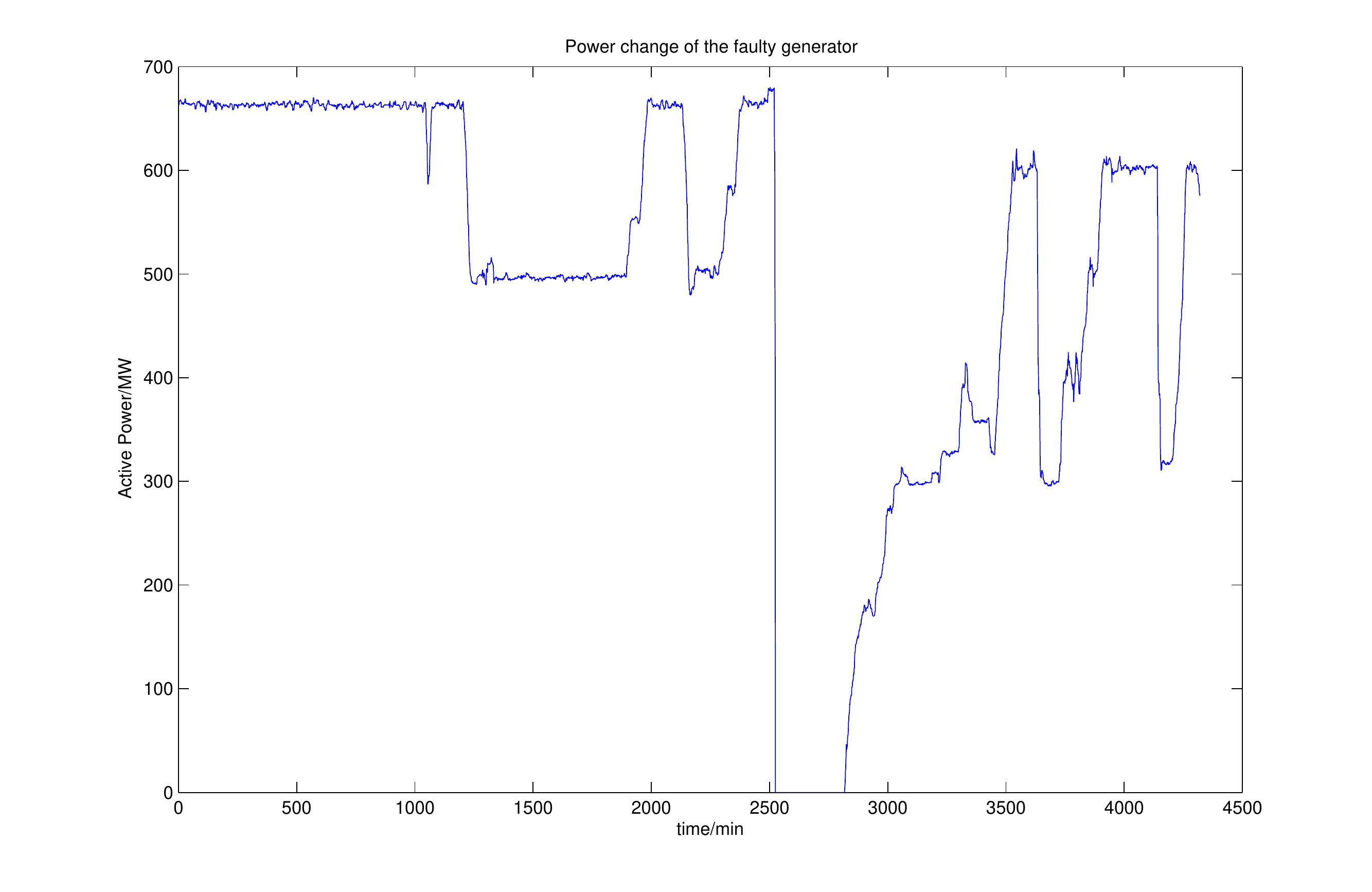}
  \caption{Power change of the faulty Generator}\label{fig:6}
\end{figure}

The mean spectral radius changes suddenly when a failure occurs as shown in Fig.4. Failure detection will be achieved through observing the relationship between the M-P Law and the empirical eigenvalue density of the matrix which is composed by the sampling data directly. Similar, the drastic change of the  mean spectral radius indicates the occurrence of failure. The result is consistent with the results in PCA method shows in Fig.5. And they are very close to the real situation shows in Fig.6. Unlike the data we get from simulation, the output of all generators are always changing in the real power grid. when one generator fails, other generators will increase their output to maintain the stability of the grid, so the mean spectral radius returns to the original position (Fig.4) earlier than when the fault is removed (Fig.6).

\section{Conclusion}
A novel random matrix method is proposed to detect power grid failure. Compared with the widely used principal components, our method exploits the convergence of the empirical spectral density, a phenomenon arising from high-dimensional probability space. We compared two methods using the real-life datasets collected in China. It seems that two methods work well in the case of our system data.
The next stage of research is to use sketching as a tool [15,16]. Our proposed sketching method projects the data set into a lower-dimensional subspace. Dimensionality reduction techniques, like e.g. principal component analysis, are commonly used in statistics. However, their focus is usually on reducing the number of variables. Our method aims to reduce the number of observations while keeping the algebraic structure of the data. This leads to a speed-up in the subsequent (frequentist or Bayesian) regression analysis, because the run-time of the common algorithms usually heavily depends on the data size.


%

\ifCLASSOPTIONcaptionsoff
  \newpage
\fi



%

%

\vfill
\vfill
\vfill



\end{document}